\documentclass[12pt]{article}
\usepackage{amsmath,cmmib57}
\usepackage{amssymb}
\usepackage{amsthm,mathrsfs}
\newcommand{\vv}{u}
\newcommand{\xx}{v}

\newcommand{\p}{\partial}
\newcommand{\Z}{\mathbb{Z}}
\newcommand{\C}{\mathbb{C}}
\newcommand{\bigmat}{\mathbb{M}}
\newcommand{\pP}{\mathbb{P}}
\newcommand{\CP}{\mathbb{C}\mathbb{P}}
\newcommand{\rd}{{\rm d}} 
\newcommand{\ri}{\rm i} 
\newcommand{\inner}[1]{{\rm i}_{#1}^{~}}
\renewcommand{\lambda}{w}
\newcommand{\cO}{\mathcal{O}}

\newcommand{\cM}{\mathcal{M}}
\newcommand{\cN}{\mathcal{N}}
\newcommand{\cP}{\mathcal{P}} 
\newtheorem{thm}{Theorem}[section]

\newtheorem{lemma}[thm]{Lemma}
\newtheorem{prop}[thm]{Proposition}

\newcommand{\oo}{\mathrm}
\newcommand\oskip{\par\addvspace{4mm}\par}

\begin{document}
\title{The geometry of dual isomonodromic deformations.}
\author{G. Sanguinetti and N.M.J Woodhouse}
\maketitle
\begin{abstract}The JMMS equations are studied using the geometry of
the spectral curve of a pair of dual systems. It is shown that the
equations can be represented as time-independent Hamiltonian flows on
a Jacobian bundle.
\end{abstract}
\par\addvspace{8mm}\par
\oskip\section{Introduction}In this paper we revisit
the geometry of dual isomonodromic deformations
of a linear system on $\C\pP_1$. The duality was originally studied,
at least in the form in which we are interested, by Harnad in
\cite{H}---although it is closely related to the `Laplace
transform' in the theory of Frobenius manifolds \cite{D},  \cite{BJL}).

The problem is to construct
isomonodromic deformations of a linear meromorphic
differential operator on the Riemann sphere $\C\pP_1$ of the form
\begin{equation}\frac{\rd}{\rd \lambda}-b 
-\sum_{i=1}^n \frac{N_i}{\lambda-a_i}\, ,\label{system1}
\end{equation}
where the $N_i$s are rank one, $r\times r$ matrices, and
$b=\oo{diag}(b_1,\ldots ,b_r)$, with the $b_i$ distinct.  The
corresponding linear system of differential equations has $n$ regular
singularities at the points $a_i$, together with an irregular
singularity of Poincar\'e rank one at infinity, where the leading
coefficient has distinct eigenvalues.

The Isomonodromic deformations  are governed by the 
JMMS equations \cite{JMMS}. These appear, or are closely related to
equations that do appear, in many areas of mathematical
physics, from impenetrable bose gases, where they were originally
introduced, to Frobenius manifolds and Seiberg-Witten problems
(see \cite{D}, \cite{Do}). 

We shall consider deformations that fix the position of the pole at
infinity, but change the parameters $b_i$ and $a_i$.  Such
deformations are equivalent to the deformations
of a second `dual system' of the same type, in which the roles of the
two sets of parameters $a_i$ and $b_i$ are interchanged \cite{H}.
The dual system  is constructed by using the rank-one
property of the $N_i$s to write the first linear operator in the form
\begin{equation}
\frac{\rd}{\rd \lambda}-G^t(a-\lambda I_n)^{-1}F +b\, ,\label{system2}
\end{equation}
where $G, F$ are $n\times r$ matrices and
$a=\oo{diag}(a_1,\ldots a_n)$. 
Then the dual operator is
\begin{equation}
\frac{\rd}{\rd z}-F(b-z I_r)^{-1}G^t +a\, .\label{system3}\
\end{equation}
At a formal level, the duality is given by writing the first linear system
of ODEs in the form
\begin{eqnarray*}
\left(\frac{\rd}{\rd \lambda}+b\right)\xx -G^t\vv&=& 0 \\
(a-\lambda)\vv+F\xx &=&0\, ,
\end{eqnarray*} 
where $\xx $ is an $r$-vector and
$\vv$ is an $n$-vector---this reduces to 
\eqref{system2} on elimination of $\vv$. 
Then the second system is given by a formal
application of the Laplace transform, to replace $\rd/\rd \lambda$ by
$z$, and $z$ by $-\rd /\rd \lambda$, as was done in a related context by
Balser, Jurkat and Lutz in \cite{BJL}
 and later in the context of
Frobenius manifolds by Dubrovin \cite{D1}.

In his original investigation, Harnad considered Hamiltonian flows on
certain subspaces of the loop algebra $\widetilde{\rm gl}(n)_-$\, ,
the subspaces being given by varying the term $F(b-z I_r)^{-1}G^t$ in
\eqref{system3}. He showed that they were generated by the polynomials
invariant by the adjoint action of the loop group $\widetilde{\rm
Gl}(n)$.  By writing down the explicit Lax equations, he deduced that
they correspond to the isomonodromic deformations of the system that
change the position of the regular singularities. He showed that they
could be viewed as time-dependent Hamiltonian flows on the symplectic
manifold of pairs of $n\times r$ matrices, with respect to a canonical
symplectic form.  The `times' are the positions of the poles, and the
Hamiltonians are 
\begin{equation}
H_i(\cN)=\frac{1}{4\pi \ri}\oint\oo{tr}
(\cN(\lambda, a))^2d\lambda\label{Hamilto}
\end{equation}
where 
$$\cN(\lambda, a)=G^T(a-\lambda I_n)^{-1}F +b$$
and the integral is around a loop containing only the $i$-th pole. They
generate the non-autonomous system
\begin{eqnarray*} 
\frac{\partial N_j}{\partial a_i}&=&\frac{{[N_j,N_i]}}{a_j-a_i},
\qquad j\neq i,\quad i,j=1,\ldots,n\\
\frac{\partial N_j}{\partial a_i}&=&
\left[b+\sum_{j\neq i}\frac{N_j}{a_j-a_i},N_i\right]\, .
\end{eqnarray*}

Here, we shall show that the Hamiltonian theory can be understood in a
different way by introducing natural variables conjugate to the $a_i$s
and $b_i$s.  We associate with the linear system \eqref{system1} a
{\em spectral curve} $\Sigma$ in $\CP_1\times \CP_1$, given by
$$\det (\cN(a,w)-z I_r)=0\, ,$$ and a line bundle $L\to \Sigma$ of
degree $nr$ (essentially the line bundle determined by the `divisor
coordinates' in \cite{AHH}).  The spectral curve and the line bundle
for the system and its dual are the same (apart from interchanging the
two copies of $\CP_1$).

The symplectic manifold $\cM$ on which the flows are defined consists
of triples $(\Sigma, B, s)$, where $B$ is a line bundle of degree zero
over $\Sigma$ and $s$ is a section defined in a neighbourhood of the
points $w=\infty$ and $z=\infty$.  Two such sections are regarded as
equivalent if the first two terms in their Taylor expansion of $s$ at
these points are the same, so the dimension of $\cM$ is $2(nr+n+r)$.

There is a standard line bundle over $\CP_1\times \CP_1$ (less
$(\infty,\infty)$) with transition function $\exp (wz)$: this encodes
the behaviour of the differential operators at their singularities.
The Hamiltonians $h_i$ are labelled by the points at infinity, and
they are in involution.  They are defined by associating a meromorphic
section of $B\otimes E$ with each point at infinity.  The zeros $q_i$
of the section are the `divisor coordinates' in \cite{AHH}, and the
Hamiltonian is given by a residue at infinity constructed from the
section. 

The two operators are recovered by treating the coordinates $w$ and
$z$ as multiplication operators on the sections of a line bundle $L$
of degree $nr$, which has divisor made up of the $q_i$s, together with
points at infinity.  Under a Hamiltonian flow, the $q_i$s move and 
$L$ is twisted over the point at infinity associated with the
Hamiltonian.  The role of $E$ here is critical: it gives rise to the
twist, which can be seen as having its origin in the evolution of the
diagonal exponentials in the formal solutions of the linear systems in
\cite{JMU}.  Harnad's non-autonomous Hamiltonian description is the
Marsden-Weinstein reduction by the Hamiltonian action.

\medbreak
\noindent {\bf Acknowledgement}.  We thank Marta Mazzocco for drawing our
attention to John Harnad's work, and John himself for interesting
discussions.

\oskip\section{The spectral curve}
The relationship between the two systems is driven by the geometry of
their common spectral curve $\Sigma\subset 
\C\pP_1\times \C\pP_1$. This has equation
\begin{equation}
\oo{det}\, \Bigl(z I_r-G^T(a-\lambda I_n)^{-1}F +b\Bigr)=0\, ,\label{realcurve}
\end{equation}
where $\lambda$ and $z$ are the coordinates on the two copies of
$\C\pP_1$. It is $n$-fold cover of the $z$-copy of $\C\pP_1$, and an
$r$-fold cover of the $\lambda$-copy.  For generic values of $F$ and
$G$, which we shall assume, it is a smooth curve.
The dual problem gives the same curve (with the two coordinates
interchanged).
 This follows either by using
the identity
\begin{eqnarray}\oo{det}(a-\lambda I_n)\oo{det}(zI_r-G^T(a-\lambda I_n)^{-1}F
+b)= \nonumber \\ \oo{det}(b-z I_r)\oo{det}(\lambda I_n-F(b-z
I_r)^{-1}G^t +a)\, ,\label{curve}\end{eqnarray}from \cite{H}; or by
noting that the curve is the set $\{(\lambda,z)\}$ on which the linear
equations
\begin{equation}
\left\{\begin{array}{c} (a-\lambda)\vv -F\xx  =0 \\ G^t\vv -(b-z)\xx 
=0\end{array}\right.\label{eveqn}
\end{equation}
have nonzero solutions for $\xx \in \C^r$, $\vv\in
\C^n$.  There are $r$ points, denoted by $x_1, \ldots ,x_r$,
at which $\lambda=\infty$, $\vv =0$, and $\xx $ is an
eigenvector of $b$; and there are $n$ points $y_1, \ldots , y_n$
at which $z=\infty$, $\xx =0$, and $\vv $ is
an eigenvector of $a$.  

The spectral curve is given for both
systems by the vanishing of
$$\oo{det}
\left[\bigmat -\begin{pmatrix}\lambda & 0\\0& z\end{pmatrix}\right]
=0\qquad \mbox{where}\qquad 
\bigmat =\begin{pmatrix}a&-F\\ -G^t&b\end{pmatrix}\, ,
$$
and the solutions to the linear equations embed $\Sigma$ in 
$\C\pP_{n+r-1}$.

\begin{prop} 
The genus of $\Sigma$ is $g=(n-1)(r-1)$.
\end{prop}

\noindent This follows by equating $2g-2$ to the number of zeros of the 1-form
$\alpha$ introduced below in eqn (\ref{1forma}); all the zeros are 
at $\lambda=\infty$ or $z=\infty$.

\oskip\section{Line bundles} The second ingredient in the construction
is the line bundle $L\to \Sigma$, defined to be the dual of the
tautological bundle on $\C\pP_{n+r-1}$, restricted to $\Sigma$. The
curve determines the eigenvalues of a matrix at each point, and the
fibres of the line bundle are dual to the corresponding
eigenspaces---that is, $L$ is dual to the line bundle given by the
solutions to the linear system \eqref{eveqn} at each point of $\Sigma$.

\begin{prop} For generic $F$ and $G$, 
$\oo{deg}(L)=nr$.\label{degr}
\end{prop}
\begin{proof} Let $\tilde F$, $\tilde G$, 
$\tilde b$ denote the matrices obtained by
deleting the first column from $F$, $G$, and the first row and column
from $b$. Let $\tilde L\to \tilde \Sigma$ be the line bundle and curve determined
by $a,\tilde b,\tilde F,\tilde G$, and put $\delta=\oo{deg}(L)$ and  
$\tilde \delta=\oo{deg}(\tilde L)$.
The map that sends 
$(\vv , \xx )$ to the first component of $\xx $ 
is a holomorphic section of $L$, and so has
$\delta$ zeros. Of these, $r-1$ are at
$\lambda=\infty$, and $n$ are at $z=\infty$. The remainder are at the finite
values of $\lambda$ and $z$ at which there
are nonzero solutions to the linear equations
\begin{eqnarray}(a-\lambda)\tilde \vv  -\tilde F\tilde \xx  =0\nonumber\\
\tilde G^t\vv  -(\tilde b-z)\tilde \xx  =0\nonumber\\
G^{1t}\tilde \vv =0\, ;\label{ind}\end{eqnarray}
in other words they are given by the finite zeros of the holomorphic section
$(\tilde \xx ,\tilde \vv )\mapsto \tilde G^{t}\tilde \vv $ of $\tilde L$. Such a section has $r-1$
zeros at $\lambda=\infty$ and is nonvanishing (for generic
$G$) at $z=\infty$. It therefore has $\tilde \delta-r+1$ at finite
values of $\lambda$ and $z$.  Hence
$$\delta-(r-1)-n=\tilde \delta -(r-1)$$
which gives
$\delta=\tilde \delta+n.$
For $r=1$, the only points of $\Sigma$ at which $\xx _1=0$
are the $n$ points above $z=\infty$, so the result 
follows by induction on $r$.

\end{proof}

\noindent 
For generic matrices, $H^0(\Sigma,L)$ has dimension $nr-g+1=n+r$, with
the global sections of $L$ determined by constant row vectors 
of length $n+r$.

Let us put
$$
L_1=L\otimes L_{y_1}^{-1}\otimes \cdots \otimes L_{y_n}^{-1}, \qquad
L_2=L\otimes L_{x_1}^{-1}\otimes \cdots \otimes L_{x_r}^{-1}\, ,
$$
and denote by $\pi_1$ and $\pi_2$ the projections from $\Sigma$ onto
the $\lambda$- and $z$-spheres. Then $L_1$ and $L_2$ have
respective degree $n(r-1)$ and $r(n-1)$, and
$$
\oo{deg}(\pi_{1*}L_i)=\oo{deg}(L_i) + 1-g -\oo{deg}(\pi_i)=0\, .$$

\begin{prop} The direct images of the bundles $L_1$ and $L_2$
on $\C\pP_1$ given by the two projections above are degree $0$
vector bundles (and hence generically trivial) of rank $r$ and
$n$ respectively.
\end{prop}

\noindent The global sections of $\pi_{1*}L_1$ and $\pi_{2*}L_2$ are given
by constant row vectors in 
$\C^r$ and $\C^n$, respectively.  In the same way as in \cite{H1},
multiplication of the corresponding sections of 
$L_1$ and $L_2$ over $\Sigma$ by $z$ and $w$, respectively, 
determines two meromorphic matrix-valued functions $Z(\lambda)$ and $W(z)$.
The original linear system
of linear differential equations and its dual are
$$
\frac{\rd \xx }{\rd \lambda}=Z(\lambda)\xx \qquad \mbox{and}\qquad
\frac{\rd \vv }{\rd z}=-W(z) \vv \, .
$$

\oskip\section{Infinitesimal deformations.}

An infinitesimal isomonodromic deformation of a linear operator of the
form \eqref{system2} is given by making an infinitesimal `singular
gauge transformation' by $\Omega(\lambda)$, where $\Omega$ is a
matrix-valued rational function of $\lambda$, with poles at the
singularities of the operator. It must be chosen so that the transformed
operator has the same singularity structure as the original \cite{JMU}.

Our starting system is equivalent to
\begin{equation}\left\{\begin{array}{c} (a-\lambda)\vv   -F\xx  =0 \\
-G^t\vv  + (b-\partial_{\lambda})\xx =0.\end{array}\right.\label{syst}\end{equation}
An infinitesimal deformation that changes the
eigenvalues at $\lambda=\infty$ will be given by $\xx \mapsto 
(1+\lambda D)\xx $ where $D=\oo{diag}(d_1,...d_r)$ is 
a small diagonal matrix. The deformed system is 
$$\left\{\begin{array}{c} (a-\lambda)\vv   -F(1+\lambda D)\xx  =0 \\
-G^t\vv  + (b-\partial_{\lambda})(1+\lambda
D)\xx =0\end{array}\right.$$ On ignoring terms of order $D^2$, the
second equation is equivalent to
$$\partial_{\lambda}\xx -b\xx +D\xx +(1-\lambda D) G^t\vv =0.$$
Now 
\begin{eqnarray*}(1-\lambda D)G^t\vv 
&=&(1-\lambda D)G^t(a-\lambda)^{-1}F(1+\lambda D)\xx \\
&=&-[D, G^tF]\xx +G^t(a-\lambda)^{-1}F\xx   \\
&&\qquad {}-DG^ta(a-\lambda)^{-1}F\xx+G^ta(a-\lambda)^{-1}FD\xx \end{eqnarray*}
where we have used
$\lambda(a-\lambda)^{-1}=a(a-\lambda)^{-1}-1$,
and ignored terms involving $D^2$.  Therefore the deformed system is
equivalent to \eqref{syst}, but with $a,b,F,G$ changed by
$$
\delta a=0, \quad \delta b=-D-[D, G^tF],\quad \delta G^t=-DG^ta, \quad\delta F=aFD.$$
Equivalently, to the first order in $D$, $\bigmat$ is deformed to
\begin{equation} 
\left(\begin{array}{cc}1 & 0 \\ DG^t &
1\end{array}\right)\left[\bigmat -\begin{pmatrix} 0& 0\\0& D\end{pmatrix}\right]
\left(\begin{array}{cc}1 & -FD \\ 0
&1\end{array}\right)\, .\label{isomon}
\end{equation}
Generally, however, the new matrix $b+\delta b$ is not diagonal.  To get back
to a system of the original form, we make a further constant gauge
transformation by $C$, where
\begin{equation}C_{ij}=\frac{(G^tF)_{ij}(d_i-d_j)}{b_i-b_j}\quad 
i\neq j\label{B}\, .\end{equation}
(Note that we require that the $b_i$s should be distinct).
The net result is to replace $\bigmat$ by
\begin{equation}
\bigmat '=
\left(\begin{array}{cc}1 & 0 \\ DG^t &
1-C\end{array}\right)\left[\bigmat -\begin{pmatrix} 0& 0\\0& D\end{pmatrix}\right]
\left(\begin{array}{cc}1 & -FD \\ 0
&1+C\end{array}\right)\, .\label{isomon2}
\end{equation}

If instead we take $\Omega=G^tD(a-\lambda)^{-1}F$, where $D$ is now a
diagonal $n\times n$ matrix, then by a similar calculation, we get
that \eqref{syst} is changed by
$$
\delta a=D+[FG^t,D], \quad \delta b=0, \quad \delta F=-DFb, \quad
\delta G^t=bG^tD\, .
$$
On rediagonalising $a$, this gives
\begin{equation}
\bigmat'=
\left(\begin{array}{cc}1+C & DF \\ 0 &
1\end{array}\right)\left[\bigmat -
\left(\begin{array}{cc} D& 0\\
0 & 0\end{array}\right)\right]\left(\begin{array}{cc}1-C & 0 \\ -G^tD
&1\end{array}\right)\, ,
\label{isomon3}\end{equation}
where now
$$
C_{ij}=\frac{(FG^t)_{ij}(d_i-d_j)}{a_i-a_j}\, .
$$
This is an isomonodromic deformation in which the positions of the
poles at $\lambda=a_i$ move to the eigenvalues of $a+\delta a$,
without changing the $b_i$s.  

By comparing \eqref{isomon2} and \eqref{isomon3}, it is clear that 
isomonodromic deformations of the original system are also
isomonodromic deformations of its dual.

\oskip\section{Elementary deformations}
\label{elem}
We say that a deformation is {\em elementary} whenever $D$ has rank one,
so that either it changes just one eigenvalue at infinity, or
moves just one of the regular singularities.  What is the effect on 
$L$ and $\Sigma$ of an elementary deformation of the form
\eqref{isomon3}, where $D_{11}=\epsilon$, with the other entries zero?

We shall denote the deformed curve and line bundle by $L'\to \Sigma'$, and use a prime to indicate the quantities associated with them. 
To the first order in $D$, we have
$$
\bigmat '-
\begin{pmatrix}\lambda &0\\0&z\end{pmatrix}\\
=
 \left(\begin{array}{cc}1+S& DF \\ 0 &
1\end{array}\right)
\left[\bigmat -\begin{pmatrix}\lambda +D&0\\0&z\end{pmatrix}\right]\\
\left(\begin{array}{cc}1 -S& 0 \\ -G^tD&1\end{array}\right)
$$
where $S=C+\lambda D$.  So we obtain $\Sigma'$ by replacing $a$
by $a-D$ in $\bigmat $ (this is true only to the first
order---that is, only for {\em infinitesimal} deformations).

Now consider the section $\sigma$ of $L$ given by $(\vv ,\xx )\mapsto
\nu(\vv )$, where $\nu$ is a constant row vector with $1$ in the first entry, 
and
zeros in the other entries.  Then $\sigma=0$ at $x_1,
\ldots,x_r,y_2,\ldots, y_n$, where $\Sigma$ and $\Sigma'$ intersect.
Denote its other zeros by $q_1, \ldots
q_{g}$. At each such point $q$, we have $D\vv =0$ and
$$
\bigmat '\begin{pmatrix}(1+C)\vv \\\xx \end{pmatrix}=\begin{pmatrix}\lambda
(1+C)\vv \\z \xx \end{pmatrix}
$$
It follows that $\Sigma$ and $\Sigma'$ also intersect at $q$, and that$$
\sigma'\, :\, (\vv ',\xx ')\to \nu (1-C)\vv '$$ is a section of $L^{\prime}$
that vanishes at each $q_i$.  We also have that $\sigma'=0$ at 
$x_1, \ldots ,x_r$.

To understand what happens to $\sigma'$ at $y_2, \ldots ,y_n$,  
we consider the
behaviour of $\vv ,\xx $ near $z=\infty$, $\lambda=a_i$ ($i\neq 1$).  By
expanding in powers of $z$, we have
\begin{eqnarray*}
\vv &=&\vv _0+z^{-1}\vv _1+O(z^{-2})\, ,\\ 
\xx &=&z^{-1}\xx _1 +O(z^{-2})\, ,\\
\lambda&=&a_i+z^{-1}c_i+O(z^{-2})\, ,
\end{eqnarray*}
where we take $\vv _0$ to have a $1$ in the $i$th entry, with the
other entries zero.  On substituting into
\eqref{eveqn}, we obtain
$$
a\vv _1-F\xx _1=a_i\vv _1+c_i\vv _0, \qquad G^t\vv _0=\xx _1\, ,
$$
and hence that
$$
(a-a_iI)\vv _1=(H+c_iI)\vv _0\, 
$$
where $H=FG^t$.  Since $\nu(\vv _0)=0$, 
it follows that $\nu(\vv _1)=H_{1i}/(a_1-a_i)$.

Now take $z^{-1}=\epsilon$.  Since $H'=H+O(\epsilon)$ and $a'=a+O(\epsilon)$,
we have 
$$
\nu(\vv ')=\nu\bigl((1-C)\vv '\bigr)=\nu\bigl((1-C)\vv _0\bigr) + \frac{\epsilon 
H_{1i}}{a_1-a_i}=O(\epsilon^2)\, .
$$
We can conclude that the $n-1$ zeros of $\sigma$ at $y_2,\ldots,y_n$ are
shifted to zeros of $\sigma'$ at the nearby  points on $\Sigma'$
at which $z=1/\epsilon$.  Therefore, to the first order in $\epsilon$,
$\sigma'/(1-\epsilon z)$ is a meromorphic section of $L'$ with the following 
properties.
\begin{itemize}
\item It has zeros at $q_1, \ldots
q_{g}$, at $x_1, \ldots ,x_r$, and at $y_2, \ldots , y_n$.
at all of which
$\Sigma$ and $\Sigma'$ intersect;

\item It has a zero at $z=\infty$, $\lambda=a_1+\epsilon$, and a pole at the
nearby point at which $z=1/\epsilon$.
\end{itemize}
Otherwise it is holomorphic and nonzero.  So we have the following.
\begin{lemma}
The line bundle $L\to \Sigma$ has a divisor contained in the intersection of
$\Sigma$ and $\Sigma'$.  The deformed bundle 
$L'$ is the bundle over $\Sigma'$ with the  
same divisor, but twisted by $\exp(1-\epsilon z)$ at $z=\infty$,
$\lambda =a_1+\epsilon$.
\label{lprime}\end{lemma}
\noindent This can be said in a simpler way. Let $V$ be an open set in $\Sigma$
made up of $n$ punctured disks, with centres at the $n$ points at
which $z=\infty$. We use $z$ as a coordinate on $V$. 

Suppose that we are given a divisor
$$
D=\sum k_im_i, \qquad m_i\in \Sigma, \quad k_i\in \Z\, ,
$$
together with a nonvanishing  holomorphic function $P(z)$, defined on
$V$.  Then we have a line bundle on $L_{\Sigma}\to \Sigma$, 
given by tensoring $L_D$ by the bundle 
with transition function $P$.  If
we have similiar nearby objects $D'$ (a divisor on $\Sigma'$)
and $P'$, then we
can characterize the change 
from $L_{\Sigma}$ to $L_{\Sigma'}$ by
giving the change $\delta D=\sum k_i\delta m_i$
in $D$ and the change $\delta P$ in $P$ as a
function of $z$.  Of course there is a lot of redundancy because different
choices of $D$ and $P$ will give the same bundle $L_{\Sigma}$.

For each $\Sigma\in \cP$, let $E$ be
the degree-zero line bundle with transition function $P=\exp(\lambda
z)$, where $\lambda$ is defined as a function of $z$ by restriction to
$\Sigma$.  Then $L\otimes E$ is determined by $P$ and the divisor
\begin{equation}
D=q_1+\cdots +q_g + x_1+\cdots +x_r+y_2+\cdots +y_n\, .
\label{Ddefn}\end{equation}
The lemma can be restated as follows. 

\begin{lemma} The deformation of $L\otimes E$ is given by 
$\delta D=0$, $\delta P=0$.
\label{isomlemma}\end{lemma}

\noindent Lemma \ref{lprime} is also true for an elementary deformation which
changes $b_1$, except that the twist is by $\exp(1- \epsilon \lambda)$
at $\lambda=\infty$, $z=b_1+\epsilon$.

\oskip\section{Symplectic approach.}

The data of a dual pair of linear operators are encoded in the following:
\begin{itemize}
\item a curve $\Sigma$ of genus $(n-1)(r-1)$, the spectral
curve of the system;
\item an embedding of $\Sigma$ into $\C\pP_1\times\C\pP_1$, with the 
projections onto the two $\C\pP_1$s of  degree $r$ and $n$ respectively; and
\item a line bundle $L\rightarrow\Sigma$ of degree $nr$. 
\end{itemize}
We shall consider how to generate isomonodromic deformations (of both
systems) from Hamiltonian flows on a symplectic manifold $\cM$
constructed, following ideas in \cite{AHH},  from curves in
$\C\pP_1\times \C\pP_1$.

Let $M=\C\pP_1\times \C\pP_1\setminus (\infty,\infty)$  (our spectral
curves do not pass through the excluded point, and are therefore
embedded in $M$). Let $\omega=\rd \lambda \wedge \rd z$: this is a
meromorphic 2-form on $M$, and it is holomorphic and symplectic except
where $\lambda=\infty$ or $z=\infty$.  

Let $\cP$ denote the space of curves in $M$ with the
properties above (in fact we shall consider only local deformations,
so $\cP$ should be thought of as an open neighbourhood of a given
spectral  curve).  Then $\oo{dim}(\cP)=nr+n+r$. This can be seen by
representing $\Sigma\in \cP$ as the zero set of a polynomial
$p(\lambda,z)$
of degree $n$ in $\lambda$ and $r$ in $z$.  The polynomial is
determined by $\Sigma$ up to scale. Alternatively, if $Z$ is 
a local section of
$TM\vert_{\Sigma}$, then the restriction of $\inner Z \omega$ to $\Sigma$ is
a meromorphic section of the canonical bundle with poles of order at
most two at $\lambda=\infty$ and $z=\infty$. It  vanishes
whenever $Z$ is tangent to $\Sigma$. Therefore the normal bundle is
$$
N=K\otimes
\pi_1^{*}(\cO(2))\otimes \pi_2^{*}(\cO(2))\, ,$$
which has degree $2g-2+2n+2r= g-1+nr+n+r$.  

We remark that
\begin{equation}\alpha=\frac{\rd\lambda}{\partial p/\partial
z}=-\frac{\rd z}{\partial
p/\partial\lambda}\, \label{1forma}\end{equation}
is a natural holomorphic 1-form on $\Sigma$, with zeros only at the
points at infinity; by considering the orders of these, we can
compute the genus of $\Sigma$.

The manifold $\cM$ is obtained from $\cP$ by attaching an `extended
Jacobian' to each curve.  A point of $\cM$ is a point on the Jacobian
of one of the curves, together with a frame for the line bundle
defined up to the second order at each point at infinity, modulo an
overall scale.  In other words, a point of $\cM$ is represented by a triple
$(\Sigma,B,s)$ where
\begin{itemize}
\item $\Sigma\in \cP$;
\item $B\rightarrow\Sigma$ is a degree zero line bundle;
\item $s$ is a nonvanishing 
section of $B$ on a neighbourhood of the $n+r$ points
of $\Sigma$ at which $\omega$ is singular.
\end{itemize}
Two such triples $(\Sigma,B,s)$ and $(\Sigma',B',s')$ determine the
same point of $\cM$ whenever $\Sigma=\Sigma'$, $B=B'$ and $s- k s'$
has zeros of order 2 at each of the $n+r$ points at infinity, for some
constant $k$. 
The data in the frames add $2(n+r)-1$ extra dimensions, so
$$\oo{dim}(\cM)= \oo{dim}(\cP)+g+2(n+r)-1=2(nr+n+r)\, . $$

The holomorphic symplectic structure $\Omega$ on $\cM$ is analogous to
that on a cotangent bundle, with the extended Jacobians playing the
role of the fibres. The symplectic form is the exterior derivative of
a `canonical 1-form' $\Theta$, which is defined as follows.

Let $T$ be a tangent vector to $\cM$ at $(\Sigma,B,s)$ and let $Z$ be
its projection into $\cP$. That is, $Z$ is a section of $N$, so the
restriction of $\inner{Z}\omega$ to $\Sigma$ is a well-defined
meromorphic 1-from, with poles at the points at infinity.  Suppose
that $s$ is defined on an open set $V$ (not necessarily connected) 
containing the points at infinity.
Let $U\subset \Sigma$ be a second open set in the complement of the points at
infinity such that $V$ and $U$ cover $\Sigma$; and let
$\beta$ be a meromorphic section of $B\vert_U$ with equal numbers of 
poles $p_i$ and zeros $q_i$, none of which are at infinity.  We define 
\begin{equation}
\inner{T}\Theta=
\sum\int_{q_i}^{p_i}\inner{Z}\omega-\frac{1}{2\pi i}\sum\oint
\oo{log}\left(\frac{\beta}{s}\right)\, \inner{Z}\omega\, ,
\label{Theta}\end{equation}
where the integrals on the right are around contours in $V\cap U$ 
surrounding the points at infinity.
Given $\beta$, $\inner{T}\Theta$ is well defined up to the addition of terms
of the form
\begin{equation}
\oint\inner{Z}\omega=\inner{T}\rd\left(\oint\theta\right)
\label{freedom}\end{equation}
where $\theta=\lambda\, \rd z$ and the integrals are around a closed contour in
$U$.  We could for example take $\beta$ to be a meromorphic section,
so that $B=\sum (q-p)$; but for a general choice, 
$\beta$ might be highly singular at infinity.

If $\beta$ is replaced by $\beta'=m\beta$, where $m$ is meromorphic
on $U$ with zeros at the poles of $\beta$, then the $\inner{Z}\Theta$
is unchanged up to the freedom above.  This follows by applying the
following with $\gamma=\inner{Z}\omega$.

\begin{lemma}  Let $U\subset \Sigma$ be a connected open set 
with boundary $\p U$ made up of
closed contours.  Let $m$ be a meromorphic function on $U$ with
equal number of zeros and poles; and
let $\gamma$ be a holomorphic 1-form on $U$.  Then
$\log(m)$ can be defined on $\p U$ and, modulo integral multiples of the periods of $\gamma$,
$$
\frac{1}{2\pi\ri}\oint_{\p U} \log(m)\, \gamma= \sum \int\gamma\, ,
$$
where on the right, the sum is over pairs of poles and zeros of $m$, 
and the integrals are along paths in $U$ from the zero to
the pole in each pair.
\end{lemma}
\noindent By the `periods of $\gamma$', we mean the integrals of $\gamma$ around closed
contours in $U$.
The proof is by extending $\log(m)$ to the complement of a set of cuts
along closed paths on $\Sigma$ and along paths 
joining paired poles and zeros of $m$.
  
It follows that $\Omega=\rd \Theta$ is a well-defined 2-form on $\cM$.
It is given explicitly as follows.  We choose $\beta$ at each
$m\in\cM$ in a neighbourhood of a given point.  Then the points of
$\cM$ can be labelled by $\Sigma$, the zeros and poles $q_i$ and $p_i$
of $\beta$, and the function $\beta/s$ defined in an annulus around
each point at infinity. We use the coordinate $w$ to identify the
annuli around $w=\infty$ on neighbouring curves; and $z$ for those
around $z=\infty$. Then a tangent to $\cM$ is represented by a tangent
vector $Z$ to $M$ at each of the points $q$, $p$, a vector field, also
denoted by $Z$, connecting $\Sigma$ to the nearby curve in a
neighbourhood of each point at infinity, and the variation
in $\log (\beta/s)$, as a function of $w$ or $z$.
\begin{prop}The 2-form  $\Omega=d\Theta$ on $\cM$ is a closed
nondegenerate 2-form given by the following expression
\begin{equation}\Omega(T,T')=\sum_{p}\omega_p(Z,Z')-
\sum_{q}\omega_q(Z,Z')
+\sum_{\lambda,z=\infty}
\oint{\Bigl(g'\, \inner{Z}\omega -g\, \inner{Z'}\omega\Bigr)}\, .
\label{Omega}\end{equation}
\end{prop}
\noindent Note that the right-hand side vanishes identically whenever $Z=0$ and
$g$ is constant, so $\Omega$ is well defined on $\cM$ (it descends
under the quotient 
by constant scaling of $s$).

\section{The Hamiltonian system}
We shall construct a Hamiltonian on $\cM$ for each point at infinity
on $\Sigma$ which generates an isomonodromic deformations of Harnad's
dual systems.  First we deal with the Hamiltonian associated with $y_1$.

For each $(\Sigma,B,s)$, we  choose a square root of $B\otimes K$. The choices
are parametrized by $H^1(\Sigma,\Z_2)$, and can be made continuously
as $\Sigma$ and $B$ vary.  In general, since its degree is $g-1$, the
line bundle $(B\otimes K)^{1/2}\otimes E^{-1}$ has a section $\tau$
which is holomorphic except for a simple pole at $y_1$.  This section
is unique up to scale.  Denote the zeros of $\tau$ by $q_1, \ldots ,
q_g$ and denote by $\mu$ the meromorphic $1$-form on $\Sigma$ which
has zeros at the points $q_i$ and a double pole at $y_1$.  Again with
the qualification `in general', this exists and is unique up to scale,
which we fix by requiring that
$$
\mu -\rd z
$$
should be holomorphic (note that the residue of $\mu$ at $y_1$ necessarily
vanishes).

The quotient $\tau^2/\mu$ is a meromorphic section of $B\otimes
E^{-2}$. It has simple poles at the points $q_i$ and is otherwise
holomorphic. From it, we obtain a section $\beta$ of $B$ (unique up to
scale) with the following properties.
\begin{itemize}
\item It is holomorphic except for simple poles at the points $q_i$,
and for an essential singularity at $y_1$, where $\exp(2\lambda z)\beta$ is
holomorphic.

\item Its zeros are the other zeros $p_1, \ldots, p_g$ of $\mu$.
\end{itemize}
We define the function $h$ on $\cM$ 
\begin{equation} h=\frac{1}{2\pi\ri}\oint_{y_1} \oo{log}(\beta/s)\, \rd z\, ,\label
{hamil}\end{equation}
where the integral is around a contour surrounding $y_1$. 

We shall calculate the derivative of $h$ along a tangent $T'$ to
$\cM$.  First suppose that
$T'$ does not move $\Sigma$.  Put $H=\exp(2wz)\beta/s$. Then $H$
is holomorphic at $y_1$ and 
\begin{eqnarray*}
T'(h)&=& \inner{T'} \rd\left(\frac{1}{2\pi\ri}\oint_{y_1} \oo{log}(H)\, \rd z\right)\\
&=& \inner{T'} \rd\left(\frac{1}{2\pi\ri}\oint_{y_1} \oo{log}(H)\, \mu\right)\\
&=& \frac{1}{2\pi\ri}\sum \oint_{y_i} g' \mu\, ,
\end{eqnarray*}
where $g'$ is the change in $\log(\beta/s)$, which in this case is 
holomorphic at $z=\infty$.  If $T'$ moves $\Sigma$, but leaves
$\beta/s$ unchanged, then $T'(h)$=0. We conclude that 
the Hamiltonian vector
field $T=T_h$ generated by $h$ is given in the representation of the
previous section by taking
$$
\inner Z\omega\vert_{\Sigma} =\mu, \qquad g=0\, .
$$

How does $h$ generate isomonodromic deformations?  We associate a dual
pair with a point of $\cM$ by identifying $L\otimes E$ with 
$$
(B\otimes K)^{1/2}\otimes L_{x_1}\otimes \cdots \otimes L_{x_r}\otimes
\cdots L_{y_1}\otimes \cdots \otimes L_{y_n}\, .
$$
In the notation of \S\ref{elem}, 
$L\otimes E$ is given by the divisor $D$ in \eqref{Ddefn}
together with a transition function $P$, defined by $P^2=\beta/s$; and
the Hamiltonian flow gives $\delta D=0$, $\delta P=0$, which is
isomonodromic by Lemma \ref{isomlemma}. With this identification,
therefore, $h$ generates isomonodromic deformations of the dual pair
of linear operators determined by $L$.  The deformation is elementary;
it changes $a_1$, leaving $a_2,
\ldots ,a_n, b_1, \ldots ,b_r$ fixed.

If we relabel $h$, $F$, $\beta$, \ldots as $h_1$, $F_1$, $\beta_1$, and
so on, and use subscripts to denote the analogous quantities defined
with $a_1$ replaced by $a_i$.  The Hamiltonians $h_i$ 
generate deformations that move the other $a_i$, leaving $b_i$ fixed.

\begin{prop} The Hamiltonians $h_i$, $i=1, \ldots ,h_n$ are in
involution.
\end{prop}
\begin{proof}
We have to show that $T_i(h_j)=0$, where $T_i$ is Hamiltonian vector
field generated by $h_i$.  Now
\begin{eqnarray*}
T_i(h_j)&=&\inner{T_i}\left(\frac{1}{2\pi\ri}\oint_{y_j}\log(F_j)\,
\rd z\right)\\
&=&\inner{T_i}\left(\frac{1}{2\pi\ri}\oint_{y_j}\Bigl(\log(F_i)+\log(\beta_j/\beta_i)\, \rd z\right)\\
&=&\inner{T_i}\left(\frac{1}{2\pi\ri}\oint_{y_j}\log(\beta_j/\beta_i)\: \inner{Z_j}\omega\right)\\
&=&\inner{T_i}\left(\sum_k\frac{1}{2\pi\ri}\oint_{y_k}\log(\beta_j/\beta_i)\: \inner{Z_j}\omega\right)\\
&=&\sum_{\rm poles}\omega(Z_i,Z_j)-\sum_{\rm zeros}\omega(Z_i,Z_j)\\
&=&0\,
\end{eqnarray*}
where in the penultimate line, the sums are over the zeros and poles
of $\beta_j/\beta_i$. In going from the fourth to the fifth line, we
use the fact that $\beta_j/\beta_i$ is holomorphic at $z=\infty$, and
that the restriction of $\inner{Z_j}\omega$ to $\Sigma$ is nonsingular   
except at $y_j$.  The last line follows because either $Z_i$ or $Z_j$
vanishes at each pole and zero.
\end{proof}
\noindent  By interchanging the roles of $\lambda$ by  $-z$, and $z$ by $\lambda$, we similarly define
Hamiltonians $k_i$ ($i=1, \ldots , r$) that
generate the other isomonodromic flows; a direct extension of the
proof above shows that these are involution with each other and with
the $h_i$s.

One could recover the nonautonomous picture of Harnad's original paper 
\cite{H} by ignoring the bundle $E$. One would have to construct two sets of 
commuting Hamiltonians and perform some symplectic quotients. A nice byproduct 
of this approach is an explicit symplectic isomorphism between Harnad's space 
of $n\times r$ matrices and a symplectic quotient of $\cM$. For details, 
see \cite{S}.

\end{document}